\documentstyle[12pt]{article}

\begin{document}
\begin{titlepage}

\begin{flushright}
UAB-FT-329\\
February 1994
\end{flushright}

\vspace{\fill}

\begin{center}
        {\LARGE\bf  MATTER FIELDS IN THE LOOP REPRESENTATION  \\
\vskip.3cm
OF THE PARTITION FUNCTION }
\end{center}

\vspace{\fill}

\begin{center}
       { {\large\bf
        J. M. Aroca }
	\vskip 0.5cm
        Departament de Matem\`atiques, \\
        Escola T\`ecnica Superior d'Enginyers de Telecomunicaci\'o, \\
	Universitat Polit\`ecnica de Catalunya \\
	08034 Barcelona, Spain
	E-mail: JOMAROCA@mat.upc.es
        \ \\
\vspace{ 7 mm}
        and
\vspace{ 7 mm}
        \ \\
        {\large\bf  H. Fort }
	\vskip 0.5cm
        Grup de F\'\i sica Te\`orica\\
        and\\
        Institut de F\'\i sica d'Altes Energies\\
        Universitat Aut\`onoma de Barcelona\\
        08193 Bellaterra (Barcelona) Spain}
	E-mail: FORT@IFAE.es
\end{center}

\vspace{\fill}

\begin{abstract}
We present the extension of the Lagrangian $loop$ representation
in such a way to introduce matter fields. The
partition function of lattice compact U(1) Gauge-Higgs model
is expressed as a sum over closed as much as open surfaces.
These surfaces correspond to world sheets
of loop-like pure electric flux excitations and open
electric flux tubes carrying matter fields at their ends.
There is a duality transformation between this description in terms
of loop world sheets and the $topological$ representation
in terms of world sheets of Nielsen-Olesen strings
both closed and open joining pairs of monopoles.
\end{abstract}

\end{titlepage}

	In a previous paper \cite{af} we showed a natural
procedure in order to set up the {\bf Lagrangian $loop$}
representation correlative to the original Hamiltonian
$loop$ formulation \cite{gt}. In reference \cite{af},
starting with the
Villain form of the action, was obtained straightforwardly
the partition function of 4D lattice pure QED as a sum of
closed world sheets of electric loops.
In the present paper we shall extend this Lagrangian loop
approach in such a way to include matter fields.
We consider the lattice compact U(1) Gauge-Higgs model which describes
the interaction of a compact gauge field $\theta_{\mu}$ with
the scalar field $\phi = |\phi| e^{i\varphi }$.
The self-interaction of the scalar field is given by the potential
$\lambda (|\phi|^2 - |\phi_0^2)^2$. For simplicity we shall consider
the limit $\lambda \rightarrow \infty$ which freezes the radial
degree of freedom of the Higgs field (it is known that
the numerical results obtained already at
$\lambda=1$ are indistinguishable from the frozen case). Thus the
dynamical variable is compact, i.e. $\varphi \in (-\pi ,\pi)$.
This model is known to possess three phases, namely confining, Higgs
and Coulomb \cite{j}. The Higgs phase splits into a region
where magnetic flux can penetrate in form of vortices
(Nielsen-Olesen strings) and a region where the magnetic flux
is completely expelled \cite{dh}, the relativistic version of Meissner
effect in superconductivity.
Relying on this, we call this two subregions: Higgs I and II
in analogy with superconducting materials.

The Villain action of this lattice model is given by

\begin{equation}
Z = \int (d\theta ) \sum_{ n }\int (d\varphi ) \sum_{ l } \exp
(-\frac{\beta}{2}\mid\mid \nabla \theta -2\pi n\mid\mid^2
-\frac{\kappa}{2}\mid\mid \nabla \varphi -2\pi l -\theta \mid\mid^2)
\label{eq:Villain}
\end{equation}

where we use the notations of the calculus of differential forms
on the lattice of \cite {g}. In the above expression:
$\beta=\frac{1}{e^2}$, $\theta$ is a real compact 1-form
defined in each link of the lattice and $\varphi$ is a real compact
0-form defined on the sites of the lattice,
$\nabla$ is the co-boundary operator,
$n$ are integer 2-forms defined
at the lattice plaquettes, and $l$ integer 1-forms,
and $\mid\mid . \mid\mid^2 = <.,.>$.

If we use the Poisson sumation formula
$\sum_n f(n) = \sum_s \int_{-\infty}^{\infty} d\phi f(\phi)
e^{2\pi i\phi s}$ for each of the compact variables,
the partition function (\ref{eq:Villain}) transforms into

\begin{eqnarray}
Z =  \sum_{ s } \sum_{ t } \int (d\theta ) \int (d\varphi )
\int_{-\infty}^{+\infty} (d\psi )
\int_{-\infty}^{+\infty} (d\chi)
\exp(-\frac{\beta}{2}
\mid\mid  \nabla \theta -2\pi \psi \mid\mid^2   \nonumber \\
-\frac{\kappa}{2}
\mid\mid \nabla \varphi -2\pi \chi -\theta \mid\mid^2 )
e^{i<s,\psi >} e^{i<l,\chi >}.
\label{eq:Villain1}
\end{eqnarray}

Integrating in the $\psi$ and $\chi$ variables

\begin{eqnarray}
Z \propto  \sum_{ s } \sum_{ t } \int (d\theta ) \int (d\varphi )
\exp(-\frac{1}{2\beta }\mid\mid  s \mid\mid^2
-\frac{1}{2\kappa}\mid\mid \ t \mid\mid^2)  \nonumber \\
\times  e^{i<s,\nabla \theta>} e^{i<t,\nabla \varphi>}
\label{eq:V1}
\end{eqnarray}

using the partial integration rule $<\psi ,\nabla \phi>
=<\partial \psi ,\phi>$
($\partial = *\nabla*$ is the boundary operator which
maps k-forms into
(k-1)-forms ) and integrating over the compact $\varphi$ and
$\theta$ we get
the constrains $\delta (\partial t = 0)$ and $\delta (\partial s = t)$
and thus, we finally arrive to

\vspace{3 mm}


\begin{equation}
Z \propto \sum_{ s } \exp[-\frac{1}{2\beta }
 \mid\mid  s \mid\mid^2 -\frac{1}{2\kappa }
\mid\mid \partial s \mid\mid^2 ]
\label{eq:LOOP}
\end{equation}

or

\begin{eqnarray}
Z \propto
  \sum_{ s } \exp [  -\frac{1}{2\beta }
  < s ,\frac{\nabla \partial + M^2}{M^2} s> ]
\label{eq:LOOP1}
\end{eqnarray}

\vspace{3 mm}

where $M^2=\frac{\kappa }{\beta }$ is the mass acquired by the
gauge field due to the Higgs mechanism.
If we consider the intersection of one of the surfaces
defined by the integer 2-forms $s$ (open and closed surfaces)
with a t=constant plane we get pure electric loop
as much as `electromesons' configurations.
Thus, we have arrived to an expression of the partition
function in terms of the world
sheets of string-like configurations: the $loop$ (Lagrangian)
representation. In this representation the matter fields
are naturally introduced by means of open surfaces which correspond
to the world sheets of open paths, i.e. `meson-like' configurations.
The corresponding Hamiltonian description in terms of gauge invariant
path-dependent operators is the so-called
$P$-representation \cite{fg},\cite{ggt}. The creation operators
of loop-states\footnote{We still use the
term `loop' for the configurations in presence of matter fields
albeit in a relaxed sense which covers both closed as much as
open paths.} are the $\Phi(C)$
for pure gauge excitations and
$\Phi(P_x^y)$ for `meson-like'
configurations, that is

\begin{equation}
\Phi (C)|0>=\prod_{l\in C)} {\bf U}(l)|0>=|C>
\end{equation}

\begin{equation}
\Phi(P_x^y)|0>={\bf \phi}^{\dagger}(x) {\bf U}(P_x^y)
{\bf \phi}(y)|0>=|P_x^y>
\end{equation}

where $|0>$ is the zero loop state (strong coupling vacuum of the
system), ${\bf U}(l)$ are the
lattice gauge group operators,
${\bf \phi}(x)$ are the matter field operators and ${\bf U}(P_x^y)$
corresponds to the product of the $U(l)$ along the path $P$ with
ends $x$ and $y$.

\vspace{0.3cm}

Another equivalent description of the Villain form
is the $topological$ or $BKT$ (for Berezinskii-Kosterlitz-Thouless)
representation in terms of the topological objects. As our
model has two compact variables we have two topological excitations:
monopoles and Nielsen-Olesen strings \cite{bpppw}.
The $BKT$ expression for the partition function of
compact scalar $QED$ is obtained via the `$Banks-Kogut-Myerson$'
transformation \cite{bkm} (see Appendix) and is given by


\begin{eqnarray}
Z \propto \sum_{n(m)}
\exp
[-2{\pi }^2 \beta <*n(m) ,\frac{M^2}{\partial \nabla + M^2} *n(m)>]
\label{eq:BKT}
\end{eqnarray}

where the $*$ denotes forms
on the dual lattice,  $m=\partial *n$ are closed integer
1-forms attached to links which represent monopole-loops
and $*n(m)=*n-\partial *q$ are integer 2-forms attached
to plaquettes corresponding to the world sheets of both
Dirac and Nielsen-Olesen strings (with monopole loops as borders).
Thus, comparing (\ref{eq:LOOP1}) and (\ref{eq:BKT})
we can observe a complete parallelism: in both representations
we have a sum over
surfaces, and intersecting with a plane t=constant
we get closed as much as open strings with point charges at
their ends. In the first case this string-like excitations
are `electric' whilst in the second they are `magnetic'.
$Loop$ and $topological$ representations are connected by
a duality transformation.
While loops provide the most natural description of the strong
coupling confining phase, Nielsen-Olesen vortices are the relevant
excitations in the weak Higgs II sector of the phase diagram.
We want to remark that there is a slight difference between
both equivalent descriptions. In the
$BKT$ representation monopoles occur
at the ends of both the Nielsen-Olesen strings
(physical excitations) and the Dirac strings
(non physical gauge variant objects) so we have the
corresponding two types of world sheets
mixed in the 2-form $*n(m)$ of equation (\ref{eq:BKT}). On the
other hand the gauge invariant $loop$ description is simpler
and completely transparent from the geometrical point of view.

\vspace{0.5cm}

In summary, we have shown that the partition function of
U(1) Gauge-Higgs theory can be represented as the sum over
world sheets of loops (open and closed). A correspondence
between the $loop$
and the $BKT$ descriptions is patent and suggests some kind of
dual connection between the confining and the Higgs II
regions of
the phase diagram.

The next step will be the implementation of the Pauli exclusion
principle in the Lagrangian loop approach in order to include
fermionic fields. This task has been accomplished in the context
of the Hamiltonian $loop$ formalism in reference \cite{fg} where
a transparent geometrical description of `full' QED was given.

\vspace{1cm}

We wish to thank R.Gambini for
enlightening discussions and comments.

\vspace{1cm}

\appendix{\bf Appendix}

To obtain the monopole representation (\ref{eq:BKT}) we start
with the Villain
form (\ref{eq:Villain}) and fix the gauge $\nabla \varphi =0$.
Then, we parameterize
the $n=\nabla q + \bar{n}(v)$, where $q$ run over arbitrary
1-forms and $v$ over all co-closed 3-forms ($\nabla v =0$).
$\bar{n}(v)$ is a solution of $\nabla n=v$.
If we perform a translation $l \longrightarrow l+q$ we get
the expression

\begin{eqnarray}
Z = \int_{\infty}^{+\infty} (dA) \sum_{q}\sum_{
                \begin{array} {c} v\\
                              \left(  \nabla v = 0 \right)
                \end{array}
}\sum_{ l } \exp
(-\frac{\beta}{2}\mid\mid \nabla A+2\pi \bar{n}\mid\mid^2 \nonumber \\
-\frac{\kappa}{2}\mid\mid A + 2\pi l \mid\mid^2)
\label{eq:Villain_nc}
\end{eqnarray}

where $A=\theta + 2\pi q$ is a non-compact variable.
By shifting
$A$ by $2\pi l$ we find dependence on the combination
$\bar{n}+\nabla l$, which turns to be a solution of
$\nabla n=v$ so we can eliminate the $l$ variable and perform the
gaussian integration yielding

\begin{eqnarray}
Z = \sum_{
                \begin{array} {c} v\\
                              \left(  \nabla v = 0 \right)
                \end{array}
	} \exp
[-2\pi ^2 \beta <\bar{n}(v),\frac{M^2}{\nabla \partial + M^2}
\bar{n}(v)>]
\label{eq:Monopoles}
\end{eqnarray}

Performing a duality transformation we get (\ref{eq:BKT})
where $m=*v$ are
now integer closed 1-forms ($\partial m= 0$) and
$*n(m)=*\bar{n}(v)$ so $\partial *n=m$.
It is possible to express (\ref{eq:Monopoles}) in terms of
the $l$ instead of the $n$ variables which reflects the
presence of Dirac and Nielsen-Olesen sheets.

\end{document}